\documentclass[twocolumn,showpacs,amsmath,amssymb,prl,superscriptaddress]{revtex4}

\usepackage{graphicx}
\usepackage{color}

\newcommand{\degree}{\ensuremath{^{\circ}}\,\,}

\begin{document}

\title{$\alpha$-RuCl$_3$: a Spin-Orbit Assisted Mott Insulator on a Honeycomb
    Lattice}

\author{K. W. Plumb}
\author{J. P. Clancy}
\author{L. J. Sandilands}
\author{V. Vijay Shankar}
\affiliation{Department of Physics and Center for Quantum Materials , University of Toronto, 60 St.~George St., Toronto, Ontario, M5S 1A7, Canada}

\author{Y.F. Hu}
\affiliation{Canadian Light Source, Saskatoon, Saskatchewan, S7N 0X4, Canada}

\author{K. S. Burch}
\affiliation{Department of Physics and Center for Quantum Materials , University of Toronto, 60 St.~George St., Toronto, Ontario, M5S 1A7, Canada}
\affiliation{Department of Physics, Boston College, Chestnut Hill, Massachusetts 02467, USA}
\author{Hae-Young~Kee}
\affiliation{Department of Physics and Center for Quantum Materials, University
    of Toronto, 60 St.~George St., Toronto, Ontario, M5S 1A7, Canada}
\affiliation{Canadian Institute for Advanced Research, Toronto, Ontario, M5G 1Z8, Canada}
\author{Young-June~Kim}
\email{yjkim@physics.utoronto.ca} \affiliation{Department of Physics and Center
    for Quantum Materials, University of Toronto, 60 St.~George St., Toronto,
    Ontario, M5S 1A7, Canada}

\date{\today}

\begin{abstract}
We examine the role of spin-orbit coupling in the electronic structure of
$\alpha$-RuCl$_3$, in which Ru ions in 4d$^5$ configuration form a honeycomb
lattice. The measured optical spectra exhibit an optical gap of 220~meV
and transitions within the $\rm t_{2g}$ orbitals. The spectra can be described
very well with first-principles electronic structure calculations obtained by
taking into account both spin-orbit coupling and electron correlations.
Furthermore, our x-ray absorption spectroscopy measurements at the Ru L edges
exhibit distinct spectral features associated with the presence of substantial
spin-orbit coupling, as well as an anomalously large branching ratio. We
propose that $\alpha$-RuCl$_3$ is a spin-orbit assisted Mott insulator, and
that the bond-dependent Kitaev interaction may be relevant for this compound.
\end{abstract}

\pacs{75.10.Jm, 71.20.Be, 71.70.Ej, 78.70.Dm}

\maketitle
Novel electronic ground states can often result from the interplay of many
competing energy scales. In magnetic materials containing 4d and 5d transition
metals, the combination of electronic correlations and spin-orbit coupling
(SOC) can give rise to exotic topological phases \cite{Kitaev2006, Wan2011,
    Okamoto2007, Lawler2008, Shitade2009, Jackeli2009, Chaloupka2010,
    Pesin2010, Witczak2011, Reuther2011,Trousselet2011}. When a transition
metal ion is subject to an octahedral crystal field environment, SOC mixes the
wave functions of the triply-degenerate t$_{2g}$ electronic states and the low
energy magnetic degrees of freedom are described by spin-orbital mixed Kramers
doublets, termed $J_{\rm eff}$ states \cite{Jackeli2009, Chaloupka2010}. One of
many interesting consequences of $J_{\rm eff}$ states in real materials is the
presence of an unusual bond-dependent exchange term called the Kitaev
interaction.  This bond-dependent interaction is a crucial ingredient for
realizing a quantum spin liquid phase on a honeycomb lattice
\cite{Kitaev2006,Chaloupka2010,Crawford1994}. Thus far, large efforts have been
directed towards studying the 5d $\rm A_2IrO_3$ (A=Na or Li) compounds where
$\rm IrO_6$ octahedra share edges to form a honeycomb network
\cite{Singh2010,Liu2011, Singh2012, Choi2012, Ye2012, Comin2012,
    Gretarsson2013}.  The edge-sharing geometry suppresses isotropic Heisenberg
interactions while Kitaev interactions are believed to be substantial
\cite{Jackeli2009,Chaloupka2010}.  However, due to monoclinic and trigonal
distortions, the applicability of the localized $J_{\rm eff}$ picture to these
compounds is still controversial \cite{Mazin2012,Mazin2013}. In light of this
complication it would be extremely useful to search for a system which is free
of these distortions in which to study spin-orbit driven physics.

The 4d counterpart of iridate physics can be found in Ru$^{3+}$ (4d$^5$)
compounds. Even though the absolute value of SOC in 4d systems is smaller than
that of 5d elements, the spin-orbital mixed state may still be realized as long
as the t$_{2g}$ states remain degenerate in the absence of SOC \cite{Chen2010}.
$\alpha$-RuCl$_3$ is an insulating transition metal halide with honeycomb
layers composed of nearly ideal edge-sharing $\rm RuCl_6$ octahedra. While
earlier transport measurements have implicated $\alpha$-RuCl$_3$ to be a
conventional semiconductor \cite{Binotto1971}, subsequent spectroscopic
investigations suggest that it may be a Mott insulator \cite{Pollini1996}.
Owing to the near ideal edge-sharing honeycomb geometry and the insulating
behaviour, $\alpha$-RuCl$_3$ is potentially an excellent candidate material in
which to realize Kitaev physics. However, the microscopic origin of such an
insulating state in $\alpha$-RuCl$_3$ remains poorly understood and a
systematic investigation of the role of SOC in this material has not been
conducted until now.

\begin{figure}
\includegraphics[width=\columnwidth]{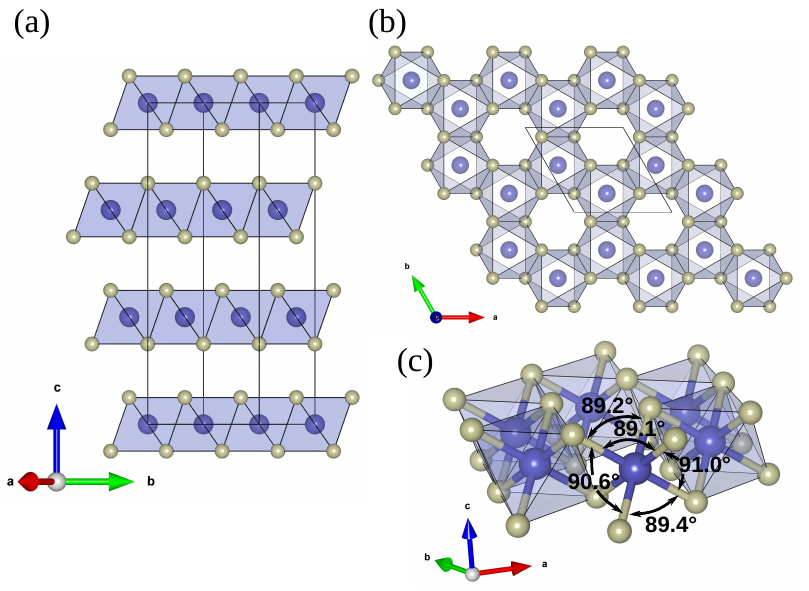}
\caption{\label{fig:structure} (Color online) (a) The crystal structure of
    $\alpha$-RuCl$_3$, exhibiting lamellar nature of the unit cell. (b)
    Individual honeycomb layers are formed by edge-sharing $\rm RuCl_6$
    octahedra (Ru in blue, Cl in grey). (c) Detailed view of $\rm RuCl_6$
    octahedra showing bond angles. All the figures were produced with VESTA \cite{VESTA}.}
\end{figure}

In this Letter, we show that the insulating state in $\alpha$-RuCl$_3$ arises
from the combined effects of electronic correlations and strong SOC\@. In order
to probe the detailed electronic structure of $\alpha$-RuCl$_3$, we have
carried out optical spectroscopy measurements. The origins of the optical gap
in $\alpha$-RuCl$_3$ are elucidated by our band structure calculations.  We
find that while strong electronic correlations are necessary to describe this
material, SOC is essential to account for the magnitude of the optical gap.
Furthermore, we have performed x-ray absorption spectroscopy (XAS) measurements
which directly indicate substantial SOC of 4d electrons. Taken as a whole, our
results indicate that $\alpha$-RuCl$_3$ is best described as a spin-orbit
assisted Mott insulator and strong SOC effects must be considered to understand
this material.

The crystal structure of $\alpha$-RuCl$_3$ is shown in
Fig.~\ref{fig:structure}.  Edge sharing RuCl$_6$ octahedra form a honeycomb
network in the a-b plane and the weakly coupled honeycomb layers are stacked
along the c-direction to form a CrCl$_3$ type structure $P3_112$
\cite{Stroganov1957}. As shown in Fig.~\ref{fig:structure} (c), the Cl-Ru-Cl
angles are all within 1\degree of 90\degree and the Ru-Cl bond lengths are
within 0.3\% of one another. Thus, the RuCl$_6$ octahedron in this compound is
very close to ideal.  In fact, the absence of appreciable electric quadrupole
interactions from the $^{99}$Ru M\"{o}ssbauer spectroscopy study was
interpreted to result from the highly symmetric octahedral configuration of the
ligand Cl ions \cite{Kobayashi1992}.  This structural detail is quite important
since such an ideal octahedral environment will leave the $t_{2g}$ states
degenerate in the absence of SOC\@.  In contrast, $\rm Na_2IrO_3$ has an O-Ir-O
bond angle of about 85\degree \cite{Ye2012,Choi2012}.  Another important
structural difference between $\rm Na_2IrO_3$  and $\alpha$-RuCl$_3$ is the
lack of intervening Na atoms between the honeycomb layers in the latter
compound, such that $\alpha$-RuCl$_3$ is closer to an ideal two-dimensional
system.

Single crystal samples of $\alpha$-RuCl$_3$ were prepared by vacuum sublimation
from commercial $\rm RuCl_3$ powder. The dielectric function
$\hat{\epsilon}(\omega) = \epsilon_1(\omega)+\epsilon_2(\omega)$ of RuCl$_3$
was measured  from 0.1 to 6~eV\@; for the range 0.9 to 6~eV,
$\hat{\epsilon}(\omega)$ was determined using spectroscopic ellipsometry. From
0.1 to 1.2~eV, we measured the transmittance through a thin RuCl$_3$ sample and
extracted $\hat{\epsilon}(\omega)$ using a standard model for the transmittance
of a plate sample \cite{reffit}. X-ray absorption spectroscopy measurements
were performed using the Soft x-ray Microcharacterization Beamline (SXRMB) at
the Canadian Light Source. Measurements were carried out at the Ru L$_3$
(2p$_{3/2}\rm \rightarrow 4d$) and L$_2$ (2p$_{1/2}\rm \rightarrow 4d$)
absorption edges.  More details of the experimental procedure are contained in
the Supplemental Material.

Physical properties of $\alpha$-RuCl$_3$ have been extensively investigated.
The magnetic susceptibility of $\alpha$-RuCl$_3$ shows a sharp cusp around
13-15~K, which was attributed to antiferromagnetic ordering
\cite{Fletcher1967}; and a Curie-Weiss fit yields effective local moment of
about 2.2~$\mu_B$ and ferromagnetic Curie-Weiss temperature of 23-40~K
\cite{Fletcher1967,Kobayashi1992}. The effective magnetic moment is much larger
than the spin only value of 1.73~$\mu_B$ for the low spin (S=1/2) state of
Ru$^{3+}$, indicating a significant orbital contribution to total moment.
Based on these observations, it was suggested that the nearest neighbor
interaction within the honeycomb plane is ferromagnetic and that these planes
are weakly coupled with an antiferromagnetic interaction.  However, powder
neutron diffraction failed to observe magnetic Bragg peaks of (003) type, which
are expected from the predicted simple magnetic structure \cite{Fletcher1967}.
Although several spectroscopic and transport investigations have been carried out to
study the electronic structure of $\alpha$-RuCl$_3$ \cite{Binotto1971,
    Guizzetti1979, Rojas1983, Pollini1996}, the role of SOC was not explored in
detail in these earlier studies.

\begin{figure}
 \includegraphics[width=\columnwidth]{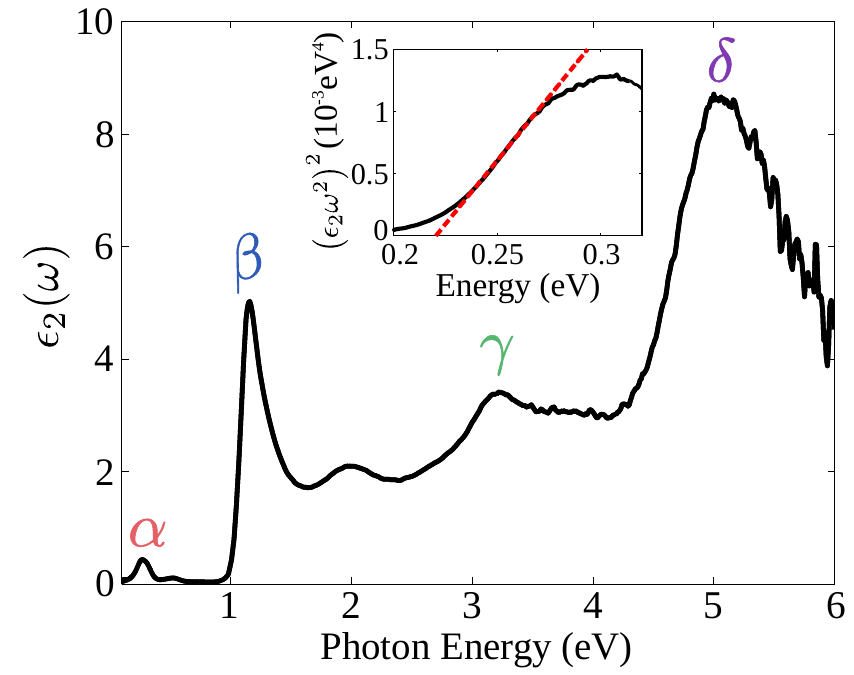}
 \caption{\label{fig:optics} Imaginary component of the dielectric function
     $\epsilon_2(\omega)$ of RuCl$_3$ measured at 295~K. The spectrum displays
     three types of excitations: transitions between t$_{2g}$ states in the
     region from 0 to 1~eV; t$_{2g} \rightarrow$ e$_g$ transitions spanning 1
 to 4~eV; and charge transfer excitations in the range of 4 to 6~eV\@.  The
 peak locations and intensities, as well as the optical gap size, are in good
 agreement with the LDA+SOC+U band structure. The transitions corresponding to
 the features labelled $\alpha$, $\beta$, and $\delta$ are shown in
 figure~\ref{fig:BandCalc} (a).  Inset: $(\epsilon_2 \omega^2)^2$ vs.\ photon
 energy in region I; the linear onset indicates an optical gap of $\approx$
 220~meV.}%
\end{figure}
In order to better understand the insulating behavior of $\alpha$-RuCl$_3$, we
have conducted optical spectroscopy measurements. In Fig.~\ref{fig:optics} we
show the measured imaginary component of the dielectric function,
$\epsilon_2(\omega)$. We find no evidence of free carrier absorption which
confirms the insulating character of RuCl$_3$. The spectrum can be divided into
three regions: i) a series of weak transitions in the range 0.1 to 1 eV, ii)
three stronger features located near 1.2, 2 and 3.2 eV, and iii) an intense
band centered near 5 eV, in agreement with previous reports
\cite{Binotto1971,Guizzetti1979}. Representative features are labeled $\alpha$, $\beta$ and $\delta$ as shown in
figure~\ref{fig:optics} to facilitate a comparison with the band structure calculation. Based on a linear
onset in the quantity $(\epsilon_2 \omega^2)^2$, shown in the inset of
Fig.~\ref{fig:optics}, we can also identify an optical gap of roughly 220~meV
at 295~K \cite{yu2010fundamentals}.

\begin{figure}
\includegraphics[width=\columnwidth]{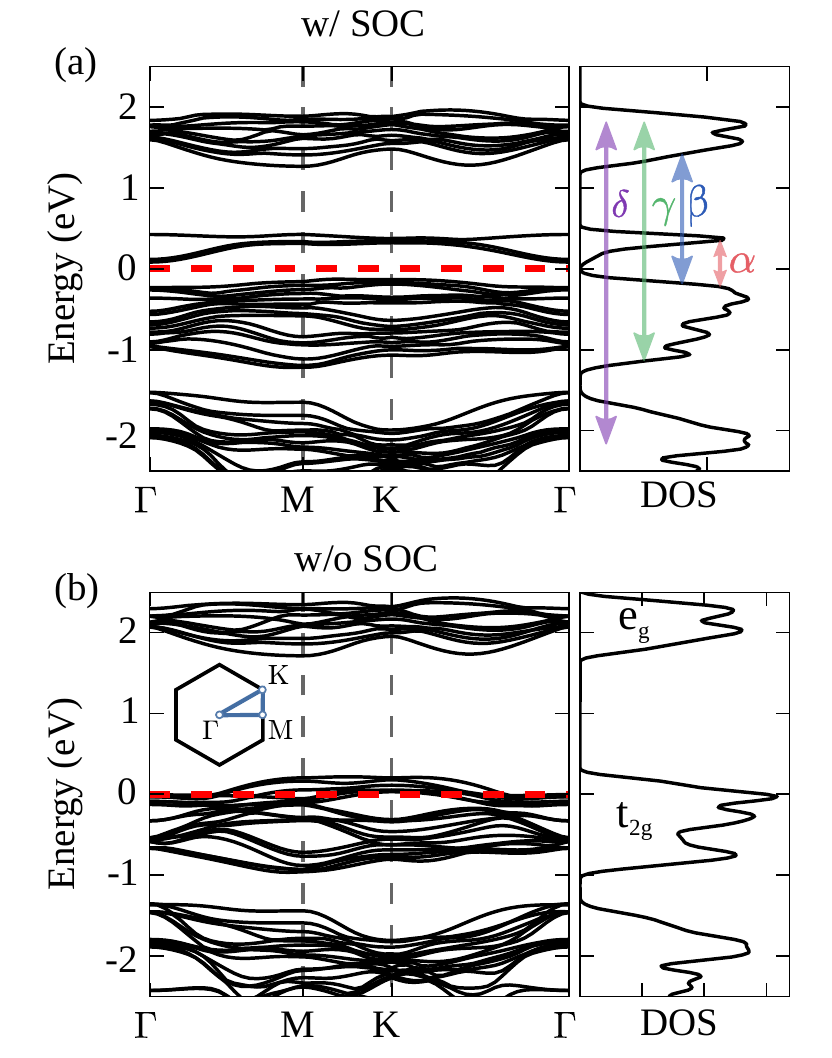}
\caption{\label{fig:BandCalc}(Color online) (a) LDA + U + SOC band structure
    and density of states (DOS) of $\alpha$-RuCl$_3$ along in plane high
    symmetry points of the BZ ($k_z\!=\!0$)\@ with U=1.5~eV and $J_H$=0.3~eV.
    Top panel is obtained with SOC and the bottom panel is without the SOC\@.
    Optical transitions denoted with arrows and labels using the same notation
    as in Fig.~\ref{fig:optics}.}
\end{figure}

The role of electronic correlations and SOC in generating the optical spectra
can be understood from our electronic structure calculations. The band
structure and total density of states (DOS) for $\alpha$-RuCl$_3$ were obtained
by performing first principles calculations including SOC and are plotted in
Fig.~\ref{fig:BandCalc}.  Details of the calculation can be found the
Supplemental Material. In Fig.~\ref{fig:BandCalc} (a), we show the band
structure and DOS obtained with Hubbard U = 1.5~eV and Hund's coupling
$J_H$=0.3~eV in the presence of SOC\@. The strength of electron correlation U =
1.5~eV was determined by comparing the direct charge gap with the measured optical gap.  The
Hund's coupling was chosen to be about 20\% of U, which is typical for 3d or 4d
transition metal compounds. On the other hand, Fig.~\ref{fig:BandCalc} (b)
presents the case with the same U and J$_H$ strengths as in
Fig.~\ref{fig:BandCalc}(a), but in the absence of the SOC\@.  For both cases,
one can see clearly the t$_{2g}$ and e$_g$ crystal field splitting due to the
octahedral environment.  However, the key difference is that
Fig.~\ref{fig:BandCalc}(a) shows an insulating phase with an unambiguous charge
gap, while the band structure is metallic when the SOC is absent as shown in
Fig.~\ref{fig:BandCalc}(b). To obtain an insulating state without SOC, a
Hubbard U value greater than 2.5~eV is required. This in turn produces a much
larger value for the charge gap which is constrained by the measured optical gap.
Therefore, a reasonable description of the insulating phase in
$\alpha$-RuCl$_3$ is only possible through the combination of SOC and electron
correlation.

Our LDA+U+SOC band structure also agrees well with the optical spectra at higher energies.
The $\alpha$ peak, together with the other weak features below 1 eV, can be understood as transitions between $t_{2g}$ states. We assign the $\beta$ feature to the lowest energetically allowed transition between the
$t_{2g}$ and $e_g$ states as represented by the arrow in Fig.~\ref{fig:BandCalc}(a); the features at 2 and 3.2 eV (labelled $\gamma$) also involve this
combination of initial and final states.  Finally, we interpret the strong peak
near 5~eV (feature $\delta$) as due to transitions from the band 2~eV below
the Fermi level to the $e_g$ states.  Indeed, our DFT calculations suggests the
band at -2~eV has an increased Cl~$p$ content, meaning the $\delta$ transition
has a charge transfer character. Overall, our optical spectroscopy measurements and electronic structure
calculations agree well, and thus identify $\alpha$-RuCl$_3$ as a spin-orbit assisted Mott
insulator.

\begin{figure}
\includegraphics[width=\columnwidth]{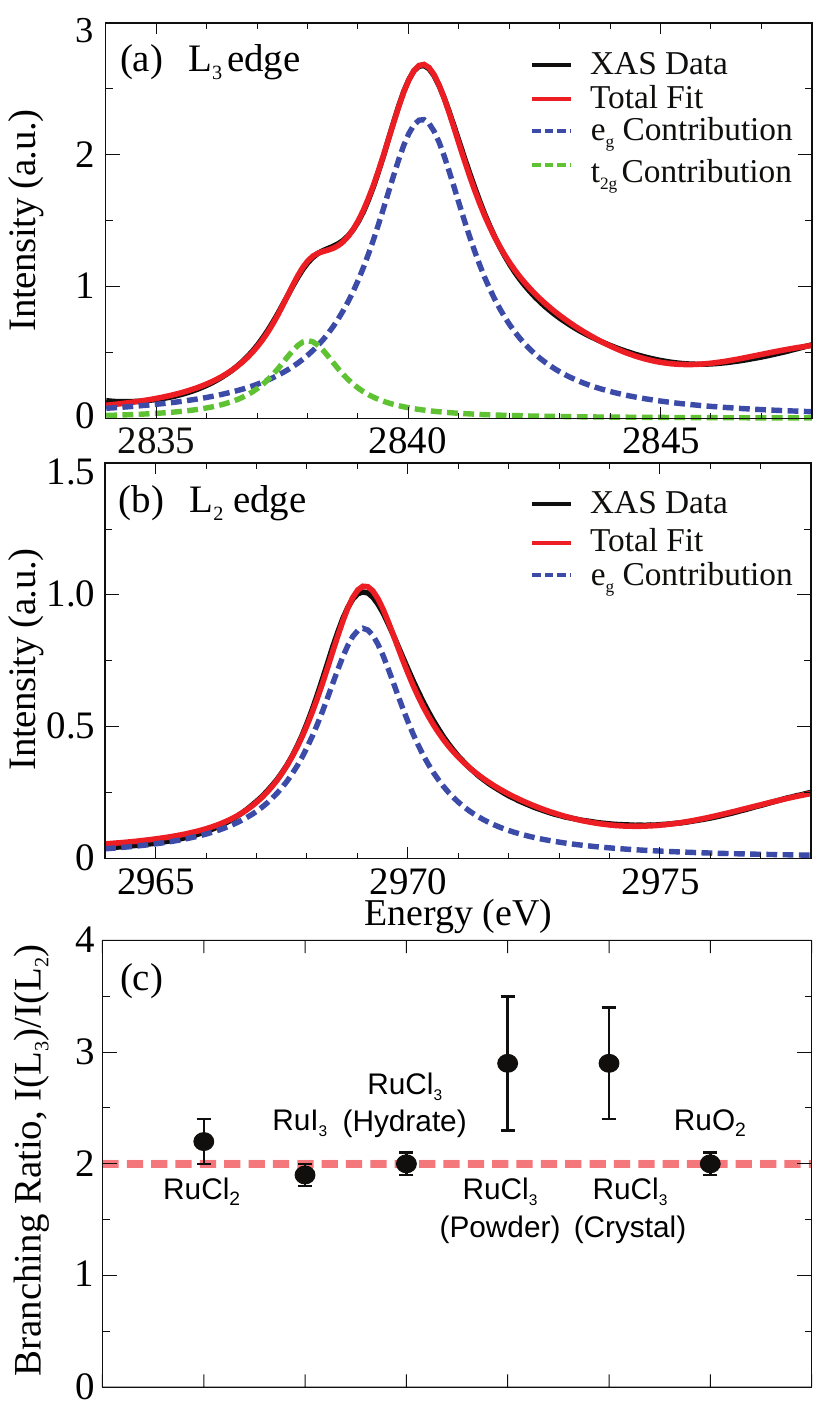}
\caption{\label{fig:XAS}(Color online) (a) X-ray absorption near edge spectra
    of RuCl$_3$ measured at the Ru L$_3$ edge. The black solid line is the
    experimental data, and the red solid line is a fit function that includes
    two Lorentzian peaks associated with $t_{2g}$ and $e_g$ states and an
    arctan function describing the edge jump. (b) Same spectra showing the
    energy range of the Ru L$_2$ edge. The scale is exactly half of the one
    shown in (a), emphasizing the departure from statistical branching ratio of
    2. (c) Comparison of the branching ratio with various Ru standard
    compounds, ranging from Ru$^{2+}$ (RuCl$_2$), Ru$^{3+}$ (RuI$_3$), to
    Ru$^{4+}$ (RuO$_2$). Note that RuCl$_3$ (hydrate) has a structure different
    from $\alpha$-RuCl$_3$ studied here.}
\end{figure}
We have independently confirmed the importance of SOC in the electronic
structure of $\alpha$-RuCl$_3$ through XAS measurements. The x-ray absorption
spectra obtained at the Ru L$_2$ and L$_3$ edges are shown in
Fig.~\ref{fig:XAS}. Two peaks are observed for the L$_3$ edge data shown in
Fig.~\ref{fig:XAS} (a), corresponding to exciting 2p$_{3/2}$ core electron into
empty t$_{2g}$ and e$_g$ states. The intensity ratio between these two features
is related to the fact that there is only one empty t$_{2g}$ state available
for the transition compared to four empty e$_g$ states. A quantitative
description of the intensity and the peak splitting requires ligand field
multiplet calculations and is beyond the scope of this letter. Here we instead
focus on the different lineshapes observed near the Ru L$_2$ edge compared to
that of the L$_3$ edge. In particular, the lower energy shoulder corresponding
to the transition to the t$_{2g}$ state is absent for the L$_2$ edge data. The
different lineshapes arise from SOC in the 4d electronic states.  At the L$_2$
($\rm 2p_{1/2}$) edge, the atomic dipole transition $\rm 2p_{1/2} \rightarrow
4d_{3/2}$ is allowed, while the $J$ selection rule forbids the $\rm 2p_{1/2}
\rightarrow 4d_{5/2}$ transition.  This is different from the L$_3$ edge case,
in which both $\rm 2p_{3/2} \rightarrow 4d_{3/2}$ and $\rm 2p_{3/2} \rightarrow
4d_{5/2}$ transitions are dipole allowed.  The absence of the L$_2$ peak
indicates that the empty t$_{2g}$ state takes on $J=5/2$ character; a result of
significant SOC effects.  The fact that the lineshape depends crucially on the
4d SOC was first noted by Sham \emph{et al.}\@ in their study of $\rm
Ru(NH_3)_6Cl_6$ \cite{Sham1983}, and later confirmed quantitatively in the
multiplet calculation carried out by de~Groot \emph{et al.}\@
\cite{deGroot1994}.

Another quantity often used to illustrate the strength of SOC is the so-called
branching ratio, defined as the main peak (`white line') intensity ratio
between the L$_3$ and L$_2$ absorption features. Typically, this value is about
two. However, when the d-electron SOC is significant, anomalously larger values
have been observed; for example, many iridate compounds show large branching
ratios \cite{Clancy2012}. If we take both peaks in the L$_3$ edge data into
account, the branching ratio of $\alpha$-RuCl$_3$ is also quite large: $3.0 \pm
0.5$. In Fig.~\ref{fig:XAS} (c), the observed branching ratios for several Ru
containing compounds are compared. Clearly $\alpha$-RuCl$_3$ exhibits an
anomalously large value. Thus, both the lineshape and the branching ratio
indicate that the SOC in $\alpha$-RuCl$_3$ is substantial.

The perceived similarities of both the crystal and electronic structure between
$\rm Na_2IrO_3$ and $\alpha$-RuCl$_3$ naturally raises questions regarding the
relevance of the Kitaev model to $\alpha$-RuCl$_3$. As mentioned earlier, $\rm
Na_2IrO_3$ is under intense scrutiny due to the possibility of realizing a
Kitaev spin liquid phase \cite{Kitaev2006, Shitade2009, Chaloupka2010,
    Reuther2011, Trousselet2011, Singh2010, Singh2012, Liu2011, Subhro2012,
    Choi2012, Ye2012, Comin2012, Liu2012, Gretarsson2013, Chaloupka2013}.
However,  the trigonal distortion present in $\rm Na_2IrO_3$ brings the atomic
basis of the spin-orbit coupled J$_{\rm eff}$=1/2 states into question
\cite{Mazin2012,Mazin2013}. Furthermore, Na atoms may promote non-negligible
further neighbor exchange terms additional to the nearest neighbor terms
\cite{Kimchi2011,Chaloupka2013}. $\alpha$-RuCl$_3$ is free from such complexity
as it is close to the ideal two-dimensional honeycomb lattice.  Even though the
atomic SOC is weaker, the ratio of the SOC and the electronic bandwidth is only
slightly smaller than in $\rm Na_2IrO_3$ because both are reduced in
$\alpha$-RuCl$_3$ compared to iridates. Indeed we find the bandwidth of
$\alpha$-RuCl$_3$ to be about half of that in $\rm Na_2IrO_3$, while the SOC is
smaller by a factor of $\sim$3. More detailed electronic structure calculations
have found that the bands near the Fermi level in $\alpha$-RuCl$_3$ are mostly
composed of J$_{\rm eff}$=1/2 except in the region near the $\Gamma$ point
\cite{Vijay}; this situation is similar to perovskite iridates
\cite{Carter2013a,Carter2013b}.  Another important difference between $\rm
Na_2IrO_3$ and $\alpha$-RuCl$_3$ is the large size of Cl anions which expands
the lattice; the Ru-Ru distance is about 10\% larger than the Ir-Ir distance in
$\rm Na_2IrO_3$. As a result, the direct hopping between the Ru t$_{2g}$
orbitals is suppressed, and indirect hopping through Cl, which gives rise to a
Kitaev interaction, is the most dominant hopping process in $\alpha$-RuCl$_3$.
Then a microscopic spin model relevant for $\alpha$-RuCl$_3$ should be composed
of both the nearest neighbor Heisenberg and bond-dependent exchange terms
denoted by Kitaev $K$ and $\Gamma$ \cite{Rau2014,Yamaji2014,Katukuri2014}.

In conclusion, we have carried out combined optical spectroscopy, electronic
structure calculations, and x-ray absorption spectroscopy investigation of the
role of spin-orbit coupling in $\alpha$-RuCl$_3$. We find that both spin-orbit coupling and
electron correlations are necessary to produce an electronic structure
consistent with the observed optical gap of about 220~meV.  In addition, the
calculated electronic structure agrees with measured higher energy optical
transitions. Our x-ray absorption spectra clearly illustrate that spin-orbit coupling of the 4d
electron system in this compound is significant. Thus spin-orbit coupling plays an essential
role in the microscopic magnetic Hamiltonian, and $\alpha$-RuCl$_3$ is likely
to exhibit unconventional magnetic ordering arising from bond-dependent
exchange interactions which could be investigated in future studies.

\begin{acknowledgements}
Research at the University of Toronto was supported by the NSERC,
CFI, OMRI, and Canada Research Chair program. Computations were performed on
the gpc supercomputer at the SciNet HPC Consortium \cite{loken2010scinet}.
SciNet is funded by: the Canada Foundation for Innovation under the auspices of
Compute Canada; the Government of Ontario; Ontario Research Fund - Research
Excellence; and the University of Toronto. Research described in this paper was
performed at the Canadian Light Source, which is funded by the Canada
Foundation for Innovation, the Natural Sciences and Engineering Research
Council of Canada, the National Research Council Canada, the Canadian
Institutes of Health Research, the Government of Saskatchewan, Western Economic
Diversification Canada, and the University of Saskatchewan.
\end{acknowledgements}

\end{document}